\pdfoutput=1
\documentclass[./supp]{./new_tlp} 
\usepackage{./supp}

\usepackage{graphicx}
\usepackage{url}
\usepackage{verbatim}
\usepackage{color}
\definecolor{lgray}{gray}{0.92}
\definecolor{lblue}{rgb}{0.90,0.90,1.00}
\definecolor{lyellow}{rgb}{1.00,1.00,0.70}

\usepackage{listings}
\newenvironment{codex}{\small\verbatim}{\endverbatim\normalsize}
 
\lstnewenvironment{code}
    {\lstset{}%
      \csname lst@SetFirstLabel\endcsname}
    {\csname lst@SaveFirstLabel\endcsname}
\newcommand{\BI}[0]{\begin{itemize}}
\newcommand{\EI}[0]{\end{itemize}}

\newcommand{\BE}[0]{\begin{enumerate}}
\newcommand{\EE}[0]{\end{enumerate}}

\newcommand{\BX}[0]{\begin{codex}}
\newcommand{\EX}[0]{\end{codex}}
\newcommand{\BV}[0]{\begin{verbatim}}
\newcommand{\EV}[0]{\end{verbatim}}

\def \bscale1 {0.25}
\def \bscale {0.25}

\begin{document}


\title[Interclausal Logic Variables]
{Interclausal Logic Variables}

\author[Paul Tarau and Fahmida Hamid]
         {PAUL TARAU and FAHMIDA HAMID\\
         {Department of Computer Science and Engineering}\\
         \email{tarau@cs.unt.edu}  \email{fahmidahamid@my.unt.edu}
}

\pagerange{\pageref{firstpage}--\pageref{lastpage}}
\volume{\textbf{10} (3)}
\jdate{March 2014}
\setcounter{page}{1}
\pubyear{2014}

\pubauthor{Tarau and Hamid}
\jurl{xxxxxx}
\pubdate{15 May 2014}

\submitted{February 14, 2014}
\revised{April 15, 2014}
\accepted{March 25, 2014}

\maketitle

\begin{abstract}
Unification of logic variables
instantly connects present and future observations of their value,
independently of their location in the data areas of the
runtime system. 
The paper extends this property to
``interclausal logic variables'',
an easy to implement Prolog extension 
that supports instant global
information exchanges 
without dynamic database updates.
We illustrate their usefulness with two of algorithms, {\em graph coloring} and
{\em minimum spanning tree}.
Implementations of interclausal variables 
as source-level transformations and as
abstract machine adaptations are given.
To address the need for globally visible
chained transitions of logic variables we
describe a DCG-based program transformation
that extends the functionality 
of interclausal variables. 
\end{abstract}


\section{Introduction} \label{intro}
Referred to by Einstein as ``spooky action at a distance", 
quantum entanglement is
the fact that observation of the values of a particle's 
physical attributes binds instantly 
entangled particles to identical values independently of their
physical distance.

In the field of quantum computing, entanglement plays a crucial role 
in designing new algorithms and communication mechanisms,
as well as in fine-tuning physical realizations 
of quantum computing machines \cite{pana11}.

In logic programming languages, the prototypical instance of such an
``entanglement pattern''
is unification of logic variables \cite{robinson:machine:jacm:65}. It
instantly connects present and future observations of their value,
independently of their  location in the data areas of the
runtime system.

While indulging into 
deviations from the strict entanglement analogy, thinking in terms
of it can clarify some interesting algorithms that are
part of the ``folklore'' of logic programming
since the early years of Prolog.
For instance, a simple and elegant graph coloring algorithm is derived
by using logic variables to denote colors associated to a vertex. 
Avoiding cycles in graph visiting algorithms,
solving a knight's tour
puzzle or finding a Hamiltonian circuit
in a graph have also simple declarative programs
exhibiting the entanglement analogy centered on unique bindings
to logic variables.

We will revisit a few of these algorithms while proposing some
new language constructs. The fact that they are unusually 
easy to implement in a language like
Prolog, gives us hope that they will lead to interesting
uses in everyday programming.

The paper is organized as follows. Section \ref{inter} introduces 
interclausal variables. Section \ref{impl} describes their
implementation in the Styla Prolog system and discusses some
alternative source-level and WAM-level implementations.
Section \ref{atwork} describes the use of interclausal variables in
algorithms like graph coloring and minimum spanning tree 
and their use to inject dynamic code in a program without using
assert operations.
Section \ref{other} discusses a source-level implementation
of backtrackable assumptions using Prolog's DCG transformation.
Section \ref{rel} discusses related work and section \ref{concl} concludes
the paper.

\section{Interclausal logic variables}\label{inter}
A natural extension of unification seen as an instance of the 
entanglement pattern is to apply it
to variables shared among different clauses, that we will 
call here {\em interclausal variables}.

In a given logic program we could syntactically mark such a variable \verb~X~, 
shared among clauses, as \verb!~X!, 
for example
\begin{code}
a(~X).
b(~X).
\end{code}
The execution algorithm will then be modified to share bindings between \verb!~X! occurring in the two clauses as in
\begin{codex}
?- a(10),b(V).
a(10),b(V).
V = 10.
\end{codex}
At the same time, it makes sense to trail such bindings, as one would do with ordinary logic variables. This means that a query like the following would also succeed, with
a different binding
\begin{codex}
?- a(V),b(20).
a(V),b(20).
V = 20.
\end{codex}

Therefore, the semantics of interclausal variables
is the same as passing them along in a shared compound term containing them
as arguments, or passing them directly as additional arguments to all predicates
occurring in the program.
Like in the case of ordinary logic variables,
their behavior on backtracking provides a form of memory
reuse.  
At the same time, indexing of Prolog clauses provides comparable access
to the shared variables as if they would be passed along in a 
data structure or as extra arguments.

As a result of our intended semantics,
one would also expect that the interclausal variables 
are
trailed and
reset to free after the query is answered.

\section{Interclausal Variables at Work} \label{atwork}

Interclausal variables can be used in Prolog facts representing
(possibly large) graphs as markers associated to vertices.
This assertional representation can provide scalability and memory
efficiency superior to equivalent representations as a data
structure.

\subsection{Graph Coloring with Interclausal Variables}

We will start by illustrating a use of interclausal variables on a graph coloring
program, derived from a classic example exhibiting the use of logic variables as
colors to be assigned to vertices.

First we define our colors:
\begin{code}
color(red). color(green). color(blue).
\end{code}
Next we define our vertices with interclausal variables 
\verb!~C1!..\verb!~C6!
representing the colors associated to each vertex.
\begin{code}
vertex(1,~C1). vertex(2,~C2). vertex(3,~C3).
vertex(4,~C4). vertex(5,~C5). vertex(6,~C6).
\end{code} 
The graph will be described as a set of edges connecting our vertices.
\begin{code}   
edge(1,2). edge(2,3). edge(1,3). edge(3,4). edge(4,5). 
edge(5,6). edge(4,6). edge(2,5). edge(1,6).
\end{code} 
The coloring algorithm will iterate over all edges
to color their endpoint vertices and then collect
the facts describing the colorings.
\begin{code} 
coloring(Vs):-
  E=edge(_,_),findall(E,E,Es),
  color_all(Es),
  V=vertex(_,_),findall(V,V,Vs).
\end{code} 
The iteration over all edges ensures at each step
that adjacent vertices are colored differently
\begin{code}     
color_all([]).
color_all([edge(X,Y)|Es]):-
   vertex(X,C), color(C),
   vertex(Y,D), color(D),
   \+(C=D),
   color_all(Es).
\end{code}
The algorithm will return multiple possible colorings
on backtracking, as if the colors were passed along
as additional arguments to each clause.
\begin{codex}
?- coloring(Vs).
Vs = [vertex(1,red),vertex(2,green),vertex(3,blue),
      vertex(4,red),vertex(5,blue),vertex(6,green)];
...
Vs = [vertex(1,blue),vertex(2,green),vertex(3,red),
      vertex(4,blue),vertex(5,red),vertex(6,green)].
\end{codex}
At the end, the interclausal variables are ready for being reused,
back to an unbound state:
\begin{codex}    
?- listing(vertex).
vertex(1,~C1).
...
vertex(6,~C6).      
\end{codex}
Note the mild deviation from our entanglement analogy,
given that (sound) negation as failure is used
to ensure that colors associated to neighboring vertices are distinct. 

Note also that
in ASP systems \cite{gringo} or SAT-based constraint solver extensions to 
Prolog \cite{zhou13picat} that rely on grounding, interclausal variables 
could be introduced with the same semantics,
to control combinatorial explosion that depends on the 
total number of distinct variables.

\subsection{A Minimum Spanning Tree Algorithm using Interclausal Variables}
Our next example uses interclausal variables for a variant of Kruskal's
minimum spanning tree algorithm with logic variables working
as markers for connected sets of edges that grow progressively until they
cover the graph (assuming it is connected).
It has been derived from a Prolog program 
using a data structure passed along
between clauses and
posted on Usenet by the 
author in 1992\footnote{A time when such uses of logic variables were still
waiting to be uncovered.}.

The algorithm proceeds by first sorting by cost the set of edges.
\begin{code}
mst(NbOfVertices,Edges,MinSpanTree):-
  sort(Edges,SortedEdges),
  mst0(NbOfVertices,SortedEdges,MinSpanTree).
\end{code}
Next the program explores the set of edges,
given as the second argument of the predicate {\tt mst0/3}.
At a given step, 
it calls the predicate {\tt mst1/7} which decides about
unifying or not the components {\tt C1} and {\tt C2}.
\begin{code}
mst0(1,_,[]). % no more vertices left 
mst0(N,[E|Es],T):- N>1,
  E=edge(_Cost,V1,V2),
  vertex(V1,C1), % C1,C2 are the components of V1,V2 
  vertex(V2,C2), 
  mst1(C1,C2,E,T,NewT,N,NewN),
  mst0(NewN,Es,NewT). 
\end{code}
The predicate {\tt mst1/7} checks if both endpoints
of an edge are already in an incrementally grown 
set of connected edges, in which case
it skips the edge. Otherwise, if the sets represented by
{\tt C1} and {\tt C2}
are distinct, they will be merged
by unifying the variables, adding the edge
to the minimum spanning tree and counting
the vertex as processed. Note that we are reusing here
the vertex definitions of our graph coloring
program, with colors interpreted as components.
\begin{code}   
mst1(C1,C2,_,T,T,N,N):-C1==C2.       
mst1(C1,C2,E,T,NewT,N,NewN):-C1\==C2,C1=C2, 
  % Put endpoints in the same component     
  T=[E|NewT],   % Add the the edge to the MST
  NewN is N-1.  % Count a new vertex 
\end{code}
Finally the predicate {\tt test\_mst} tries out the
algorithm on a small graph.
\begin{code}  
test_mst(MinSpanTree):- 
  Edges = [ edge(70,1,3),edge(80,3,4),edge(90,1,5),
            edge(60,2,3),edge(20,4,5),edge(30,1,4), 
            edge(40,2,5),edge(50,3,5),edge(10,1,2)      
          ],
  mst(5,Edges,MinSpanTree).    
\end{code}
Note that an answer is returned as a list of edges ordered by cost,
such that each vertex is an endpoint of at least one edge.
\begin{codex}
?- test_mst(Mst).
Mst = [edge(10,1,2),edge(20,4,5),edge(30,1,4),edge(50,3,5)]
\end{codex}


\subsection{Injecting Dynamic Code without Asserts}

When  used in a metavariable position, an interclausal variable
can provide a lightweight alternative to the
assert/retract interface to dynamic code.
In a clause like 
$a_0$\verb~:-~$a_1,\dots$\verb!~V!$\ldots,a_n$
the metavariable \verb!~V! can be bound to a Prolog terms
that gets ``injected'' in the possibly statically compiled
code of the clause.
In particular, injecting \verb!~V=fail! in a clause like
$a_0$\verb!:-~V!$\ldots,a_n$ would temporarily disable the
clause without the need to use a {\tt retract} operation.
In a different branch of the computation, one could inject
 \verb!~V=true! to enable the clause. 

\section{Implementing Interclausal Variables} \label{impl}
We will describe here a few mechanisms for adding 
support for interclausal variables to Prolog systems.

\subsection{Interclausal variables in Styla}

We have implemented {\em interclausal variables} in our
{\em Styla} Scala-based Prolog system \cite{styla}  by
taking advantage of its object oriented term structure and
its distributed unification and term copying algorithms,
designed in such a way that various subterms contribute small steps depending on their 
type. We have also used the fact that 
inheritance enables ``surgical'' overriding of the small
methods implementing these algorithms together.

First, we have created a new type {\tt EVar} for interclausal variables as an extension of
the class of ordinary logic variables {\tt Var}.
\begin{code}
package prolog.terms
class EVar() extends Var {
  override def tcopy(dict : Copier) : Term = this.ref
}
\end{code}
We made it inherit all properties of logic variables except one: behavior on copying.
The method {\tt tcopy}, instead of creating a fresh variable, simply returns the 
reference {\tt ref} of our interclausal variable. As a result, bindings of
interclausal variables are shared between calls, while their unification behavior,
including trailing for undoing bindings on backtracking, is inherited unchanged.

Styla uses Scala's {\em combinator parsing} API where only two simple modifications
were needed to process our new data type.

First, we specified the regular expression
\begin{code}
val evarToken: Parser[String] = """~[A-Z_]\w*""".r
\end{code}
defining that interclausal variables start with the \verb!~! symbol and have, otherwise,
the same token specification as the usual ones.

Next, we have ensured that the parser knows about them, by adding a rule associated
to their token type calling the method {\tt mkEVar}
\begin{code}
def mkEVar(x: String) = {
    vars.getOrElseUpdate(x, new EVar())
}
\end{code}
Finally, a small change to the {\tt toString} method marks with a ``\verb!~!''
the string representation of interclausal variables. Besides helping with debugging,
this is also  useful as Styla keeps track of
variable names in the source code and uses them
in predicates like {\tt listing/1}, 
when the source code of a predicate is displayed.

\subsection{Source-level Implementations}
Given a set of interclausal variables, one can implement them at source level
simply by adding them as extra arguments to each clause of a program. This would
ensure that a Datalog program remains a Datalog program after the transformation.
While linear, the resulting code explosion can be avoided by adding a single variable
to each clause representing a compound term, together with an {\tt arg/3}
predicate call accessing the appropriate position in it,
for each interclausal variable occurring in a given clause.

\subsection{WAM-level Implementations}
In a way similar to BinProlog's implementation of
multiple DCG streams \cite{DT97:AGNL}, the argument registers 
(represented as an array in BinProlog \cite{bp2011}) can be
extended with as many positions as needed to accommodate all
interclausal variables, to which the compiler would generate 
appropriate references in instructions
like {\tt unify\_variable} and {\tt unify\_value} \cite{aitkaci:warren:91}.
Alternatively, a heap area could be reserved for them, say at a lower
address range than  that reserved for ordinary variables,
and instructions would be generated to create them on the heap before
execution begins.

\subsection{Scoping constructs and interclausal logic variables}
Limiting the scope of interclausal variables to smaller code units
can be achieved easily in the case of a source-level implementation
by limiting their addition as extra arguments to only the clauses
of a given module.

On the other hand, in a Prolog systems that would support
{\em local clauses}, with a semantics similar to Haskell's ``{\tt where}''
construct (usable for local function definitions), one could implement
variants of interclausal variables as logic variables one or more levels up
from the point where they are used with or without copying
on new clause calls.

\section{Source-level backtrackable assumptions} \label{other}
We will overview here another, less ``pure'' instance of
the entanglement pattern that provides, at source-level, a richer
set of functionalities than interclausal or backtrackable global
variables.


A limitation of interclausal variables is that they do not allow
threading information that changes over multiple recursive
calls, for which the prototypical example is Prolog's
Definite Clause Grammar (DCG) mechanism \cite{pereira:definite:ai:80},
which has been extended to support multiple independent
chains of variables at source level \cite{edcg} or
at WAM-level \cite{TarauDF95a}.

As an application of the WAM-level implementation of
\cite{TarauDF95a}, specific to the BinProlog system,
{\em Assumption Grammars} have been introduced in \cite{DT97:AGNL}
featuring backtrackable dynamic database updates
and a mechanism allowing the programmer to chose
between copying or sharing semantics for the
assumed clauses.  

We will describe here a source level implementation of the functionality
of Assumption Grammars, by overloading the standard
DCG mechanism. As a result, it is portable to virtually all Prolog systems.

Note that predicates defined here with arity {\tt 3} should be used within clauses
defined with DCG arrow ``\verb~-->/2~'' rather than the usual clause neck ``\verb~:-~/2''.

The Assumption Grammar API is implemented as follows as source level program transformation.
   
\subsection{Setting and getting the database and the DCG tokens}   
\noindent \verb~'#<'(Xs)~ sets the DCG token list to be Xs for processing by the assumption grammar.   
\begin{code}
'#<'(Xs,_,Db-Xs):-new_assumption_db(Db).
\end{code}
\verb~'#>'(Xs)~ unifies current assumption grammar token list with \verb~Xs~.
\begin{code}
'#>'(Xs,Db-Xs,Db-Xs).
\end{code}
\verb~'#:'(X)~ matches \verb~X~ against the current DCG token the assumption grammar is working on.
\begin{code}
'#:'(X,Db-[X|Xs],Db-Xs).
\end{code}

\subsection{Adding new assumptions}
\verb~'#+'(X)~ adds ``linear'' assumption +(X) to be consumed at most once, by a \verb~'#-'~ operation.
\begin{code}
'#+'(X,Db1-Xs,Db2-Xs):-add_assumption('+'(X),Db1,Db2).
\end{code}
Note that variables occurring in a clause assumed with the \verb~'#+'~operation
are ``interclausal'' and their bindings provide a long distance
communication channel between the points where they are produced and consumed.
\verb~'#*'(X)~ adds 'intuitionistic' assumption \verb~'*'(X)~ to be used 
indefinitely by \verb~'#-'~operation.
\begin{code}
'#*'(X,Db1-Xs,Db2-Xs):-add_assumption('*'(X),Db1,Db2).
\end{code}
The semantics of these clauses is essentially the same as Prolog's dynamic database
with ``immediate update'', except that assumptions are backtrackable.
\subsection{Querying the assumptions}
\verb~'#='(X)~ unifies X with any matching existing or future \verb~'+'(X)~ linear assumptions.
\begin{code}
'#='(X,Db1-Xs,Db2-Xs):-equate_assumption('+'(X),Db1,Db2).
\end{code}
\verb~'#-'(X)~ consumes a +(X) linear assumption or matches a \verb~'*'(X)~ intuitionistic assumption.
\begin{code}
'#-'(X,Db1-Xs,Db2-Xs):-consume_assumption('+'(X),Db1,Db2).
'#-'(X,Db-Xs,Db-Xs):-match_assumption('*'(X),Db).
\end{code}
Note that this operation provides a mechanism to call either linear or intuitionistic assumptions,
except that in the later case, matching assumptions are ``consumed'' i.e; removed from the database.
\noindent
\verb~'#?'(X)~ matches \verb~'+'(X)~ or \verb~'*'(X)~ assumptions without any binding.
\begin{code}
'#?'(X,Db-Xs,Db-Xs):-match_assumption('+'(X),Db).
'#?'(X,Db-Xs,Db-Xs):-match_assumption('*'(X),Db).
\end{code}

\subsection{Auxiliary predicates}

A few auxiliary predicates implement internals of the API:
\begin{code}
new_assumption_db(Xs/Xs).

add_assumption(X,Xs/[X|Ys],Xs/Ys).

consume_assumption(X,Xs/Ys,Zs/Ys):-nonvar_select(X,Xs,Zs).

match_assumption(X,Xs/_):-nonvar_member(X0,Xs),copy_term(X0,X).

equate_assumption(X,Xs/Ys,XsZs):- \+(nonvar_member(X,Xs)),!,
  add_assumption(X,Xs/Ys,XsZs).
equate_assumption(X,Xs/Ys,Xs/Ys):-nonvar_member(X,Xs).
\end{code}
Finally, \verb~nonvar_member(X,XXs)~ and \verb~nonvar_select(X,XXs,Xs)~ are 
variants of \verb~member/2~ and \verb~select/3~ working on open ended lists.
\begin{code}
nonvar_member(X,XXs):-nonvar(XXs),XXs=[X|_].
nonvar_member(X,YXs):-nonvar(YXs),YXs=[_|Xs],nonvar_member(X,Xs).

nonvar_select(X,XXs,Xs):-nonvar(XXs),XXs=[X|Xs].
nonvar_select(X,YXs,[Y|Ys]):-nonvar(YXs),YXs=[Y|Xs],nonvar_select(X,Xs,Ys).
\end{code}

\subsection{Using Assumption Grammars}
One can use phrase/3 to test out assumption grammar components, as follows:
\begin{codex}
?- phrase(('#<'([a,b,c]),'+'(t(99)),'#*'(p(88)),'#-'(t(A)),'#-'(p(B)),
           '#:'(X),'#>'(As)),Xs,Ys).
A = 99, B = 88, X = a, As = [b, c],
Ys = [*(p(88))|_G2344]/_G2344-[b, c] .  

?- phrase(('#<'([a,b,c]),'#+'(t(99)),'#*'(p(88)),'#-'(t(A)),
           '#-'(p(B)),'#:'(X),'#>'(As)),Xs,Ys).
A = 99, B = 88, X = a, As = [b, c],
Ys = [*(p(88))|_G1161]/_G1161-[b, c] .
\end{codex}
We refer to \cite{DT97:AGNL} for various examples of their use
both for expressing concisely some Prolog algorithms and
for capturing  long distance
dependencies in natural language processing phenomena 
like anaphora resolution and agreement.

Note that one could also implement similar constructs
by combining interclausal variables storing compound terms 
in which mutable backtrackable state is
updated with built-ins like {\tt setarg/3}.


\section{Related Work}\label{rel}

The first author must confess that about 25 years ago he has thought about
and even wrote a short draft paper about interclausal logic variables that
got forgotten and lost. Being quite sure that something similar might
have popped-up over time and has made it into the logic 
programming folklore, we have
not revisited the subject until now, except for 
a footnote in \cite{padl09inter} where 
inter-clausal variables are
 mentioned as write-once global variables relating them to the
semantics of term copying. Other than that, we have not found despite
an extensive search, any reference to them or closely
related concepts.

In \cite{td94:LOPSTR}, after applying the binarization transformation \cite{Tarau93a},
multi-headed clauses are introduced, 
which give direct access to continuations at source-level.
The technique makes possible long distance communication
between logic variables otherwise inaccessible.
 
Global variables (both backtrackable and persistent)
have been present in BinProlog since the mid-1990s \cite{dbt95a} and
are these days available in various Prolog systems.
Among them, we mention {\tt SWI-Prolog}'s implementation \cite{wielemaker2012swi}
where their values live on the Prolog global stack. 
Like in the case of interclausal variables,
this implies that lookup time is independent 
of the size of the term. As a result, they can
efficiently store large data structures like
parsed XML syntax trees or 
global constraint stores.

By contrast to non-backtrackable
global variables, our interclausal variables are
single assignment and behave similarly to
ordinary logic variables.
Backtrackable global variables are semantically similar to interclausal variables.
However, like in BinProlog 2.0's original implementation \cite{Tarau94:BinProlog},
they are named with constants and used through an API like
SWI-Prolog's {\tt b\_setval/2} and {\tt b\_getval/2}, requiring a hash-table look-up
to find their values on the heap, while the interclausal variables in this proposal
are implemented simply as a special case of logic variables resulting also
in a more natural notation.

\section{Conclusion}\label{concl}

Interclausal variables extend
natural properties of the usual logic variables
to variables shared among clauses.
Given the simplicity of their implementation, for which we have outlined
a few alternative scenarios, we hope they can contribute to adding
flexibility to logic programming languages while keeping intact
their declarative flavor. 

We have also described a source-level
implementation of ``assumption grammars'' an extension to Prolog's 
DCGs that circumvents some limitations of interclausal variables.

We plan future work on implementing interclausal variables at WAM-level
and experiments with their uses in probabilistic logic programming.
We also plan to work on mechanisms  based on interclausal logic variables
that optimize the grounding phase in ASP systems and SAT-based
constraint solvers used by Prolog systems.

\bibliographystyle{./acmtrans}
\bibliography{}


\end{document}